\documentclass[prl,superscriptaddress,amsmath,amssymb,floatfix]{revtex4}
\usepackage[latin1]{inputenc}
\usepackage{bm}
\usepackage{graphicx}

\begin{document}

\title{Observation of a One-Dimensional Spin-Orbit Gap in a Quantum Wire}

\author{C. H. L. Quay}
\affiliation{Physics Department, Stanford University, Stanford, CA 94305-4060, USA}
\affiliation{Bell Labs, Alcatel Lucent, Murray Hill, NJ 07974, USA}
\affiliation{Present address: Quantronics Group, CEA Saclay, 91191 Gif-sur-Yvette, France}

\author{T. L. Hughes}
\affiliation{Physics Department, Stanford University, Stanford, CA 94305-4060, USA}
\affiliation{Physics Department, University of Illinois Urbana-Champaign, Urbana, IL 61801, USA}

\author{J. A. Sulpizio}
\affiliation{Physics Department, Stanford University, Stanford, CA 94305-4060, USA}

\author{L. N. Pfeiffer}
\affiliation{Bell Labs, Alcatel Lucent, Murray Hill, NJ 07974, USA}
\affiliation{Present address: Department of Electrical Engineering, Princeton University, Princeton, NJ 08544, USA}

\author{K. W. Baldwin}
\affiliation{Bell Labs, Alcatel Lucent, Murray Hill, NJ 07974, USA}
\affiliation{Present address: Department of Electrical Engineering, Princeton University, Princeton, NJ 08544, USA}

\author{K. W. West}
\affiliation{Bell Labs, Alcatel Lucent, Murray Hill, NJ 07974, USA}
\affiliation{Present address: Department of Electrical Engineering, Princeton University, Princeton, NJ 08544, USA}

\author{D. Goldhaber-Gordon}
\affiliation{Physics Department, Stanford University, Stanford, CA 94305-4060, USA}

\author{R. de Picciotto}
\affiliation{Bell Labs, Alcatel Lucent, Murray Hill, NJ 07974, USA}
\affiliation{Present address: B-Nano Ltd., 2 Meir Weisgal Road, Rehovot 76326, Israel}

\begin{abstract}

Understanding the flow of spins in magnetic layered structures has enabled an increase in data storage density in hard drives over the past decade of more than two orders of magnitude~\cite{fert}. Following this remarkable success, the field of `spintronics' or spin-based electronics~\cite{fert, bratkovsky, zutic} is moving beyond effects based on local spin polarisation and is turning its attention to spin-orbit interaction (SOI) effects, which hold promise for the production, detection and manipulation of spin currents, allowing coherent transmission of information within a device~\cite{fert, bratkovsky}. While SOI-induced spin transport effects have been observed in two- and three-dimensional samples, these have been subtle and elusive, often detected only indirectly in electrical transport or else with more sophisticated techniques~\cite{kato, sih-prl, konig-science, folk-bsr, chen, xia}. Here we present the first observation of a predicted `spin-orbit gap' in a one-dimensional sample, where counter-propagating spins, constituting a spin current, are accompanied by a clear signal in the easily-measured linear conductance of the system~\cite{pershin,zhang}.

\end{abstract}

\maketitle

We first introduce the class of phenomena we dub `the one-dimensional spin-orbit gap' using a simple example adapted from Ref.~\cite{pershin}, then describe our experiment in detail, and finally present a more elaborate model which captures most of the features seen in our data.

The spin-orbit interaction is a relativistic effect where a charged particle moving in an electric field experiences an effective magnetic field which couples to its spin~\cite{griffiths}. In semiconductor heterostructures, the electric field can arise as a result of either the lack of an inversion centre in the crystal (bulk inversion asymmetry, BIA) or a lack of symmetry in an external confining potential (structural inversion asymmetry, SIA) due to crystal interfaces or additional structures such as metallic gates~\cite{winkler}. The strength of the resulting effective magnetic field is proportional to both the particle's momentum and the original electric field.

\begin{figure}
\includegraphics{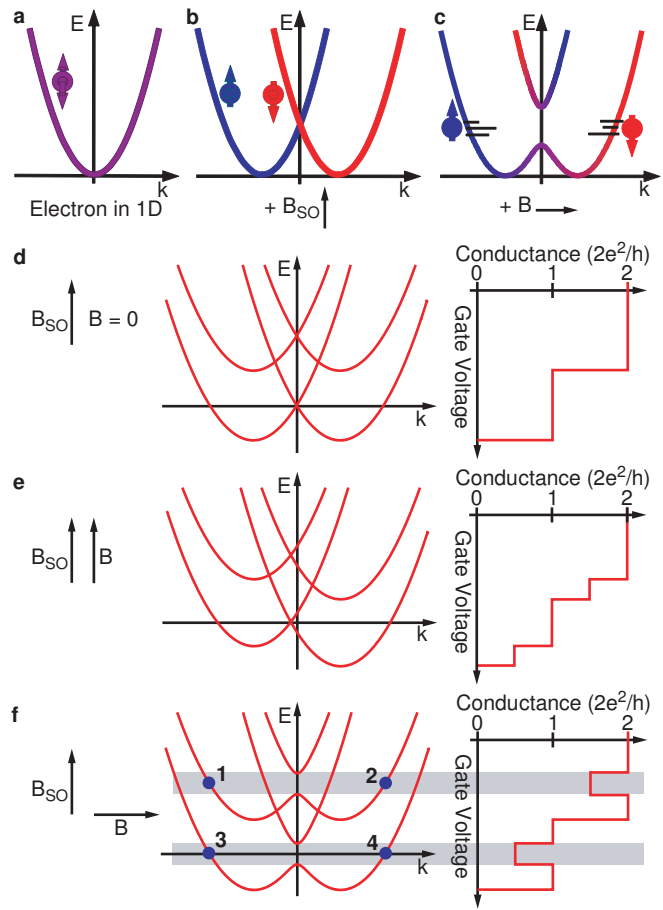}
\caption{The spin-orbit gap in a simple model and the associated conductance features. a, The dispersion relation for electrons in a spin-degenerate one-dimensional subband. b, The spin-orbit interaction (SOI) lifts the spin degeneracy, displacing spinful subbands laterally with respect to each other. c, An applied magnetic field can mix the two spin bands, creating the anti-crossing we dub the 'spin-orbit gap'. When the Fermi energy is tuned to be within this gap, particles of opposite spin travel in opposite directions, producing a spin current. At the same time, a clear drop is expected in the conductance, shown in f. d, With SOI but without an applied magnetic field, the conductance of the system increases by a step of $2e^2/h$ each time the gate voltage aligns the Fermi energy to the bottom of a subband. e, A magnetic field applied parallel to $\protect\overrightarrow{B}_{SO}$ splits each step into two half-steps of $e^2/h$. f, With a magnetic field applied perpendicular to $\protect\overrightarrow{B}_{SO}$, spin-orbit gaps appear as in c and the conductance drops when the Fermi energy lies within such a gap.}
\end{figure}

Consider a spin-degenerate one-dimensional subband with a Hamiltonian $H_0 = \frac{\hbar^2k^2}{2m}$ , where $\hbar$ is Planck's constant, $k$ the particle's  momentum and $m$ its mass (Figure 1a). The leading order SO contribution to the Hamiltonian is $H_{SO}=\beta\overrightarrow{\sigma}\cdot(\overrightarrow{k}\times\nabla V)$, where $\overrightarrow{\sigma}$  is the particle's spin, $V$ the electrostatic potential and $\beta$ a material-dependent parameter~\cite{elliott}. This term breaks the spin-degeneracy of the system and results in two spinful subbands separated by a lateral (wave vector) shift as shown in Figure 1b.

Despite this rather striking change in the band structure, measurements of conductance through the system cannot distinguish the situation shown in Figure 1b from the case where the spins are degenerate. In both cases the edges of the two spin subbands occur at the same energy, so in both cases the conductance rises by $G_0 = 2e^2/h$  when the Fermi level of the system is tuned through this energy (e.g. by applying a voltage to a nearby gate)~\cite{cronenwett-diss, kouwenhoven, glazman}. (Figure 1d.) To detect the SO shift in a transport measurement, a different approach is needed.

Note that the spins as shown in Figure 1b are polarized in the direction of $\overrightarrow{B}_{SO}$, which is perpendicular to both the momentum (i.e. the 1D wire) and the external electric field~\cite{footnote1}. A magnetic field applied exactly along $\overrightarrow{B}_{SO}$ shifts the spinful bands up and down by the Zeeman energy respectively, splitting each step of size $G_0$ into two steps of size $G_0/2$ as shown in Figure 1e. Here spin is polarized alternately up and down for the respective spinful bands and thus the charge current can be completely spin-polarized when only one spinful band is occupied; however, charge transport measurements still reveal no difference between this system and one where there is no SOI.

Now consider a magnetic field applied perpendicular to $\overrightarrow{B}_{SO}$ (say along the wire): the two spinful bands are mixed by this term so that the zero field crossing-point becomes an anti-crossing. We call this feature the `spin-orbit gap' (Figure 1c). When the Fermi level lies within such a gap, two Fermi points (e.g. 3 and 4 in Figure 1f) contribute to the current as opposed to the four at the same energy in Figure 1d. The conductance is thus reduced by $G_0/2$ as shown in Figure 2d. In addition, the current from this subband is completely spin polarized in a way that is expected to be robust to moderate disorder: holes must scatter between the points 1 and 2 or between 3 and 4 (Figure 1f) and perform a spin flip in order to backscatter. In such a system of counter-propagating spins (very nearly what is termed a `helical liquid'~\cite{wu}), the direction of the spin current is expected to be independent of the sign of the voltage applied across the system and is determined only by the sign of the SOI. Intriguingly, a pure spin current, without charge current, is expected to exist at zero bias voltage.

Finally, a magnetic field applied neither exactly perpendicular nor parallel to $\overrightarrow{B}_{SO}$ results in a mixture of the two effects described above~\cite{pershin}.

We turn now to our experiment. Our samples are GaAs/AlGaAs hole quantum wires produced by the cleaved edge overgrowth (CEO) method~\cite{loren-ceo,lorenapl}. Starting with an extremely high-mobility 2D hole gas (2DHG) realised in a carbon-doped AlGaAs/GaAs/AlGaAs quantum well~\cite{manfra-2dhg}, the sample is cleaved and additional AlGaAs is grown using Molecular Beam Epitaxy (MBE) over the freshly exposed surface. Further carbon-doping leads to hole accumulation at the GaAs/AlGaAs interface on the cleavage plane, resulting in a 1D wire (Figure 2a). We are able to apply a magnetic field either parallel or perpendicular to the wire (in the $x$ or $y$ directions indicated in Figure 2b).

The basic properties of our wires have been or will be reported in other publications~\cite{lorenapl}. Another body of work on quasi-one-dimensional wires in GaAs by Hamilton et al.~\cite{hamilton-njp,hamilton-prl-06} focused on gate-defined quantum point contacts on a 2DHG grown in the 311 direction. These studies found that, due to the SOI, the effective g-factor depends strongly on the particular subband studied as well as the direction of the wire and that of the magnetic field with respect to the crystal axes. They did not, however, observe signs of the SO gap reported below, perhaps due to the growth direction of their 2DHG or their nearly-symmetrical confinement potential.

\begin{figure}
\includegraphics{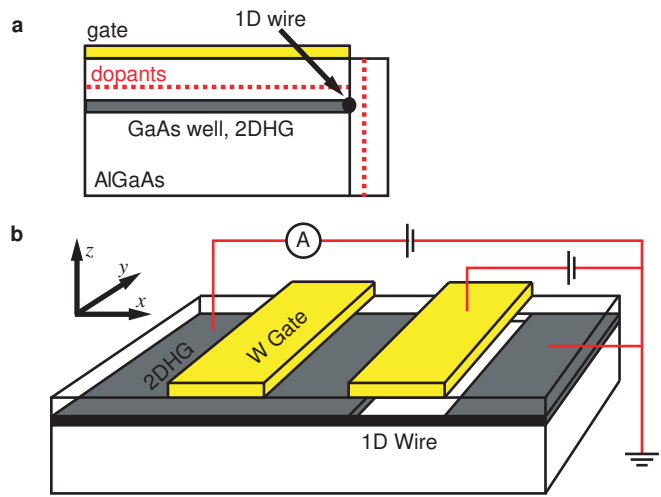}
\caption{The device and measurement setup. a, Cross-section of our devices, which are fabricated by the cleaved-edge overgrowth method~\cite{loren-ceo,lorenapl}. b, Measurement scheme for the one-dimensional hole wires. A section of the wire is isolated using a gate which depletes the two-dimensional hole gas (2DHG) just beneath it. Conductance is measured using ohmic contacts to the 2DHG on either side of the wire and decreases in steps each of which corresponds to the depletion of a 1D subband~\cite{lorenapl}}
\end{figure}

In Figure 2b, applying a positive voltage to pre-fabricated gates on the top surface of the wafer, we deplete first the 2DHG under the gate to isolate the 1D wire and subsequently the subbands of the wire. The conductance, measured between ohmic contacts to the 2DHG on either side of the gate/wire, decreases in steps as the wire subbands are successively depleted~\cite{footnote-stepheight,rafi-2d1d}.

\begin{figure}
\includegraphics{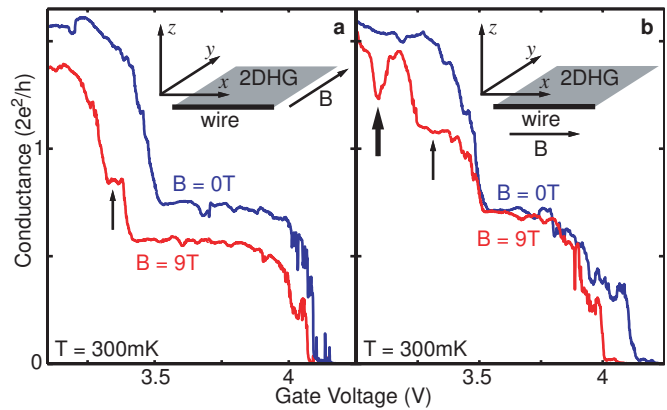}
\caption{The first two conductance steps of a quantum wire (blue traces) and their evolution in magnetic fields applied in two different directions (red traces). a, A field perpendicular to the wire splits the second step into two half-steps, the first marked with an arrow). b, A field parallel to the wire produces a half-step (arrow) and a dip (fat arrow), signifying the presence of a spin-orbit gap. On the first step (gate voltages larger than 3.5V) no dip is observed and half-steps, if they exist at all, are hardly discernible.}
\end{figure}

In Figure 3 --- which shows our main experimental results --- we focus on the first two conductance steps seen at zero magnetic field (blue traces) corresponding to the lowest two subbands in the wire, and their evolution in magnetic fields applied in two different directions, $y$ and $z$ (red traces). Concentrating first on the second, higher, conductance step, we see that a magnetic field in the $y$ direction transforms this step into two half-height steps (Figure 3a), whereas a field in $x$ produces both two half-steps and a dip (Figure 3b). In contrast, the first step appears unaffected by magnetic field in either direction.

Let us try to understand these results in terms of the simple model presented in Figure 1. Assuming that the main contribution to the spin-orbit effect is due to structural inversion asymmetry, we expect $\overrightarrow{B}_{SO}$ to be perpendicular to the wire, in the $y$-$z$ plane, as the structure is translationally invariant along the length of the wire. Thus, a magnetic field in the $x$ direction (perpendicular to $\overrightarrow{B}_{SO}$) should produce dips in the zero-field conductance steps (as in 1f), whereas one in the $y$ direction should split the each zero-field conductance step into two half-steps (as in 1e) and possibly also produce dips simultaneously.

Our data bear more than a passing resemblance to the expectations from the simple model, yet the two are significantly different in their details. A more realistic model, described below, allows us to more fully understand our results.

We use a four band Luttinger model~\cite{winkler,footnote-luttinger} and include confinement in two directions, taking into account the orientation of our wire in the GaAs crystal. This model considers the lowest four spinful subbands, with the higher subbands being ignored. The parameters in the model are: $C$, $\gamma_1$, $\gamma_2$, $\gamma_3$, $\chi$, $d_y$, $d_z$, $r_y$ and $r_z$. $C$ and the $\gamma_i $ are bulk (Al)GaAs band structure parameters which are well-established in the literature. $\chi$ describes the leakage of the 1D hole wavefunctions from the GaAs where they are nominally confined into the surrounding AlGaAs; however the model is not very sensitive to this parameter. $d_y$ and $d_z$ describe the confinement in the $y$ and $z$ directions, and $r_y$ and $r_z$ the strength of the SIA electric field in the $y$ and $z$ directions. We start by using an isotropic value of 2 for the g-factor and return to this point later. The full Hamiltonian and details of its derivation as well as values chosen for each parameter can be found in Supplementary Information.

\begin{figure}
\includegraphics{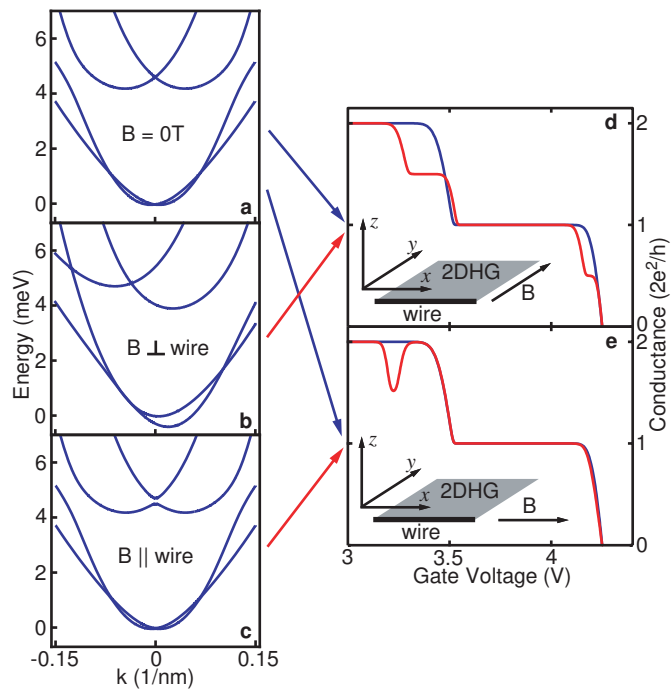}
\caption{Predictions from our model of band structures and conductances traces. a, The dispersion relations of the two lowest quantum wire subbands with no applied magnetic field. b, The same bands in a magnetic field of 9T perpendicular to the wire, with $g = 2$. c, The same bands in a magnetic field of 9T parallel to the wire, with $g = 2/9$. d, Conductance traces calculated from a (blue) and b (red) with the experiment temperature of 300mK. Compare to Figure 3a. e, Conductance traces calculated from a (blue) and c (red) with the experiment temperature of 300mK. Compare to Figure 3b.}
\end{figure}

Figure 4a shows the lowest energy bands calculated with this Hamiltonian in the absence of magnetic field. Focusing first on the upper pair of bands, we see that applying a field in the $y$ direction shifts them in energy (Figure 4b), producing a double step in the conductance trace (Figure 4d), which is seen in our data (Figure 3a). In contrast, a field in the $x$ direction produces a gap in the band structure (Figure 4c) and a dip in the conductance trace (Figure 4e), also seen in out data (Figure 3b).

As for the lower pair of bands, the spin-splitting at B = 0 is much smaller than in the upper bands. Additionally, as in our data, magnetic fields in both directions do not have much of an effect (Figures 3 and 4)~\cite{footnote-smallstep}.

Clearly the coupling between the two lower bands due to the SOI, or a magnetic field, is suppressed. This is a consequence of the fact that these bands have been chosen in the model to be primarily of heavy-hole character --- their carriers have spin $\pm 3/2$ --- a choice motivated by our data and other physical considerations as explicated in the SOM. The leading order Rashba and Zeeman terms are linear order in spin-operators and thus can only couple states with $\Delta S = 1$. As $\Delta S = 3$ for the heavy holes, the two lower two subbands are not coupled by these terms. Their coupling requires cubic order terms in the spin operators. Such terms, however, are small in all components of the Hamiltonian: Rashba and Zeeman couplings which are cubic in spin-operators do exist, but their coupling constants are suppressed~\cite{winkler}. The lowest order Dresselhaus term is cubic in spin-operators, but in GaAs it is again very small~\cite{winkler} and thus has almost no effect on the bands.

Indeed, the Dresselhaus term has a minimal effect not only on the lower bands but on the upper pair of bands as well. Therefore, the SO effects seen in our samples come primarily from the asymmetry of the confining potential of the 1D wire, i.e. the Rashba term. We find further that the electric field due to this potential is stronger in the $z$ direction than in the $y$.

We return now to the question of the g-factor. We find that, in order to produce features of the experimentally observed sizes at the applied 9T field, our model requires an effective g-factor of $~2$ in the $y$ direction and an effective g-factor of $~2/9$ in the $x$ direction. (Figure 4d and e.) Such an anisotropy in the effective g-factor is expected in the presence of SO coupling and is dependent on the strength and direction of confinement~\cite{winkler,hamilton-njp,hamilton-prl-06,footnote-gfactor}.

We are thus able to describe most of the important features of our data with the proposed model and to identify their physical causes. Further theoretical work is needed to understand one feature of our data which our model fails to capture: the occurrence of a double step in addition to a dip when the magnetic field is applied in the $x$ direction.

We also acquired data similar to those in Figure 3 at finely-spaced intermediate fields (Supplementary Information). These data show that the observed conductance dip and half-steps develop gradually with magnetic field, allowing us to rule out resonances associated with particular values of the magnetic field as the cause of the observed features. In addition, these features --- and in particular the conductance dip --- were observed in a separate device (Figure 5) and persist after thermal cycling; thus they are also not linked to particular configurations of the disorder potential in the device.

\begin{figure}
\includegraphics{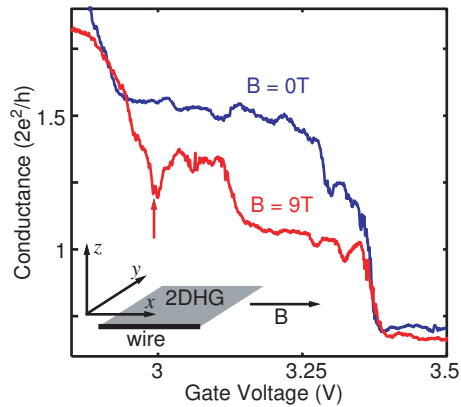}
\caption{Reproducibility. Additional data, from a separate device, showing the dip on the second step due to the presence of the spin-orbit gap.}
\end{figure}

In conclusion, we have observed for the first time the spin-orbit gap predicted for one-dimensional systems and we have developed a model which describes most of the detailed features of our data by taking into account the materials properties of (Al)GaAs.

Several directions seem promising for the further exploration of these gaps. For this particular device geometry, even cleaner wires would allow the study of SO effects at finite bias voltage and in higher subbands. A tunnelling setup such as in Ref~\cite{auslaender} could probe the dispersion relations directly; and a side gate as in Ref~\cite{yacoby-ssc} would allow tuning the strength of the electric field which gives rise to the spin-orbit effect.

For the purposes of spintronics applications, verification of spin transport, particularly at zero bias, would be of great interest. This could be achieved either through direct detection of the spin current or though the detection of spin accumulation at the two ends of the device. It would also be of practical interest to explore the possibility of producing such SO gaps in the lowest subband and without a magnetic field. While in our experiments the spin-orbit gap is induced by a magnetic field, this may not be necessary: the two spin bands could be mixed at $k = 0$ in other ways, perhaps through the use of materials with magnetic order or through controlled doping with magnetic impurities.

Finally, in other studies Coulomb interactions have been found to have marked effects in one-dimensional systems including CEO electron wires~\cite{auslaender-science}, and further work is necessary to understand their role in these newly-developed systems.

\noindent\textbf{Acknowledgements} We thank P. Joyez, M. A. Wistey, J. E. Moore and A. S. Goldhaber for helpful discussions and/or comments on the manuscript. JAS acknowledges support from a National Science Foundation graduate fellowship, CQHL support from a Harvey Fellowship and Bell Labs, and DG-G a Fellowship from the David and Lucile Packard Foundation. Work at Stanford was primarily supported by the AFOSR under contracts FA9550-04-1-0384 (PECASE) and FA9550-08-1-0427.

\noindent\textbf{Author contributions} KWW and LNP performed the MBE growth, RdP and CQHL the rest of the sample fabrication. KWB characterised the 2DHGs before the cleave as well as control samples from the overgrowth step. RdP, CQHL and JAS performed measurements. TLH derived the theoretical model with input from RdP and CQHL. CQHL, RdP, DGG, TLH and JAS analysed the data. CQHL wrote the manuscript together with RdP, DGG and TLH. RdP and LNP had the idea for the experiment.

\noindent\textbf{Competing Interests} The authors declare that they have no competing financial interests.

\noindent\textbf{Correspondence} Correspondence should be addressed to CQHL~(email: cquayhl@stanfordalumni.org).





\section{Supplementary Information}

\subsection{Derivation of the Hamiltonian and Choice of Parameters}

We begin with the four-band Luttinger model, which models the uppermost valence bands of GaAs~\cite{winkler}. This model, containing the heavy-hole (HH) and light-hole (LH) valence bands, is valid for large-gap III-V semiconductors with band edges at the $\Gamma$-point. For the bulk Hamiltonian we have

\begin{equation}
\begin{split}
&H = \cfrac{\hbar^2}{2m} \sum_k c^\dagger_k \\
&\left( \begin{matrix} \negthickspace
-(\gamma_1+\gamma_2)k^2_{\parallel}-(\gamma_1-2\gamma_2)k^2_z & 2\sqrt{3}\gamma_3k_-k_z & \sqrt{3}(\gamma_2\hat{K}-i2\gamma_3k_xk_y) &  0 \\
2\sqrt{3}\gamma_3k_+k_z & -(\gamma_1-\gamma_2)k^2_{\parallel}-(\gamma_1+2\gamma_2)k^2_z & 0 & \sqrt{3}(\gamma_2\hat{K}-i2\gamma_3k_xk_y) \\
\sqrt{3}(\gamma_2\hat{K}+i2\gamma_3k_xk_y) & 0 & -(\gamma_1-\gamma_2)k^2_{\parallel}-(\gamma_1+2\gamma_2)k^2_z &  -2\sqrt{3}\gamma_3k_-k_z  \\
0 & \sqrt{3}(\gamma_2\hat{K}+i2\gamma_3k_xk_y) & -2\sqrt{3}\gamma_3k_+k_z &  -(\gamma_1+\gamma_2)k^2_{\parallel}-(\gamma_1-2\gamma_2)k^2_z
\end{matrix} \right) \\
&c_k
\end{split}
\end{equation}

\begin{equation}
k^2_\parallel = k^2_x +k^2_y
\end{equation}

\begin{equation}
\hat{K} = k^2_x - k^2_y
\end{equation}

\begin{equation}
k_\pm = k_x \pm ik_y
\end{equation}

\begin{equation}
c_k = \begin{pmatrix}
c_{3/2k} & c_{1/2k} & c_{-1/2k} & c_{-3/2k}
\end{pmatrix}
\end{equation}

Here $c_{\sigma k}$ destroys a fermion in the $J_z = \sigma$ bulk band. As mentioned in the main article, we have found that the effects of bulk inversion asymmetry are very small and therefore inconsequential; we have therefore set the parameter $C$ in Reference~\cite{winkler} to zero to simplify the representation. The quantum well growth directions are the crystal [001] and [110] directions so we will rotate our axes around the [011] direction by an angle of -$\pi$/4. From here on the $x$, $y$ directions will be aligned along the [1$\bar{1}$0] and [110] crystal directions respectively.

In our device, holes are confined in the $y$ and $z$ directions, to create a 1D wire along $x$. Therefore, $k^2_y$ and $k^2_z$ may be replaced by their expectation values in the lowest subband: $ \langle k^2_\alpha \rangle \sim (\pi/d_\alpha)^2 $ where $d_\alpha$ is the confinement width in the $\alpha$ direction. Ignoring structural inversion asymmetry (SIA) terms, the effective 1D Hamiltonian is:

\begin{equation}
H_{1d} = \cfrac{\hbar^2}{2m}\sum_k c_k^\dagger \begin{pmatrix}
-(\gamma_1+\gamma_2)k^2_x & 0 & -i(\sqrt{3}\gamma_2k^2_x-\beta_y) & 0 \\
0 & -(\gamma_1-\gamma_2)k^2_x - \Delta & 0 & -i(\sqrt{3}\gamma_2k^2_x-\beta_y)\\
i(\sqrt{3}\gamma_2k^2_x-\beta_y) & 0 & -(\gamma_1-\gamma_2)k^2_x - \Delta & 0\\
0 & i(\sqrt{3}\gamma_2k^2_x-\beta_y) & 0 & -(\gamma_1+\gamma_2)k^2_x
\end{pmatrix}c_k
\end{equation}

with

\begin{equation}
\Delta = 4\gamma_2\pi^2\left( \cfrac{1}{d^2_z} - \cfrac{1}{2d^2_y} \right)
\end{equation}

\begin{equation}
\beta_y = \sqrt{3}\gamma_2\cfrac{\pi^2}{d^2_y}
\end{equation}

and the basis
\begin{equation}
c_k = \begin{pmatrix}
c_{0+3/2k} & c_{0+1/2k} & c_{0-1/2k} & c_{0-3/2k}
\end{pmatrix}.
\end{equation}

Here $c_{0\sigma k}$ destroys a fermion in the lowest confinement subband of the $J = \sigma$ bulk band at momentum $k$. We have chosen a pair of subbands arising from heavy-hole like states $c_{0\pm3/2k}$, and a pair of subbands arising from light-hole like states $c_{0\pm1/2k}$. Although it is not clear a priori which bulk bands would give rise the four lowest subbands, this choice is motivated by the fact that, given Equation (7), the spacing between the chosen subbands could be comparable to the inter-subband spacing of the heavy-hole-like series of subbands. Thus, the two reasonable options seemed to be either that the two lowest sub-bands should be heavy-hole-like or that one of them should be light-hole-like. Our choice of the latter is subsequently validated by our data (see main article). The parameter $\beta_y$ arises from the confinement in the $y$ direction~\cite{winkler}.

We now return to the SIA (Rashba) terms. These may be deduced from symmetry, or more formally the method of invariants~\cite{winkler}. For spin-3/2 systems there are two allowed Rashba terms which are linear in momentum:
\begin{equation}
H_{SIA}= \alpha_1\left(\overrightarrow{k} \times \overrightarrow{E} \right) \cdot \overrightarrow{J} + \alpha_2\left(\overrightarrow{k} \times \overrightarrow{E} \right) \cdot \overrightarrow{J'}
\end{equation}
with $\overrightarrow{J} = (J_x,J_y,J_z)$ the spin-3/2 matrices and $\overrightarrow{J} = (J^3_x,J^3_y,J^3_z)$. Empirically $\alpha_2 \ll \alpha_1$ so we shall ignore the second term~\cite{winkler}. For the confinement electric fields we have $E_z \neq 0$ and $E_y \neq 0$ and thus the 1D Rashba term is
\begin{equation}
\begin{split}
H^R_{1d} &= \alpha_1 k_x(E_y J_z - E_z J_y)\\
&=\begin{pmatrix}
\frac{3}{2}r_y k_x & \frac{\sqrt{6}}{4}\phi_+r_z k_x & 0 & 0 \\
\frac{\sqrt{6}}{4}\phi_-r_z k_x & \frac{1}{2}r_y k_x & \frac{\sqrt{2}}{2}\phi_+r_z k_x & 0 \\
0 & \frac{\sqrt{2}}{2}\phi_-r_z k_x  & -\frac{1}{2}r_y k_x & \frac{\sqrt{6}}{4}\phi_+r_z k_x \\
0 & 0 & \frac{\sqrt{6}}{4}\phi_-r_z k_x & -\frac{3}{2}r_y k_x
\end{pmatrix}
\end{split}
\end{equation}
where $r_y = \alpha_1 E_y$, $r_z = \alpha_1 E_z$ and $\phi_\pm = 1 \pm i$.

Finally, we consider terms added to the Hamiltonian to describe the effects of an applied external magnetic field. We shall ignore here orbital magnetic field effects because our magneto-conductance data does not reveal substantial shifts of the band edges with field, which are to be expected had orbital effects been important. As for the Zeeman terms, two are expected for the spin-3/2 systems:
\begin{equation}
H^Z= g_1\mu_B\overrightarrow{B} \cdot \overrightarrow{J} + g_2\mu_B\overrightarrow{B} \cdot \overrightarrow{J'} .
\end{equation}							
Here again $g_2 \ll g_1$ and therefore the second term can be ignored~\cite{winkler}. In matrix form, the remaining (first) term is:
\begin{equation}
H^Z = g\mu_B\begin{pmatrix}
\frac{3}{2}B_z & \frac{\sqrt{6}}{4}\phi_-B_- & 0 & 0 \\
\frac{\sqrt{6}}{4}\phi_+B_+ & \frac{1}{2}B_z &  \frac{\sqrt{2}}{2}\phi_-B_- & 0 \\
0 & \frac{\sqrt{2}}{2}\phi_+B_+ & -\frac{1}{2}B_z &  \frac{\sqrt{6}}{4}\phi_-B_- \\
0 & 0 & \frac{\sqrt{6}}{4}\phi_+B_+ & -\frac{3}{2}B_z
\end{pmatrix}
\end{equation}
with $g$ the Landé g-factor, $\mu_B$ the Bohr magneton, $B_\pm=B_x\pm iB_y$ and  $\phi_\pm=1\pm i$.

Our full model Hamiltonian is:
\begin{equation}
H = H_{1d}+H^R_{1d}+H^Z.
\end{equation}

We have calculated the resultant dispersion relations for these four low-lying bands using a sensible set of parameters listed below and deduced the magneto-conductance traces expected from this model. Despite the various assumptions and approximations made, this model captures faithfully most of the features in our data semi-quantitatively (see main article). One feature in our data is not captured by this model; however, pointing to a particular assumption or approximation made here as the primary cause of this shortcoming is beyond the scope of this manuscript.

We now turn to the values used for the various model parameters. We have used known values where possible and the reasoning for the choices made for the remaining parameters is detailed below. The values used are:

\begin{table}[ht]
\caption{Parameter values used in the model.}
\centering
\begin{tabular}{|c|c|}
\hline
Parameter & Value used in model \\
\hline
$\gamma_1$ & 6.85 (5.68)\\
$\gamma_2$ & 2.1 (1.63) \\
$\gamma_3$ & 2.9 (2.35) \\
$\chi$ & 0.05 \\
$d_y$ & 60nm \\
$d_z$ & 25nm \\
$r_y$ & 0 \\
$r_z$ & 23$\pm$2meV$\cdot$nm \\
\hline
\end{tabular}
\end{table}

The $\gamma_n$ are crystal parameters and are well-established in the literature~\cite{winkler}. The numbers in brackets are for Al$_{0.324}$Ga$_{0.676}$As calculated using a linear interpolation in the virtual crystal approximation, while those not in brackets are for GaAs. The actual numbers used in the model are $\gamma_n = \chi\gamma_{n,AlGaAs} + (1 - \chi)\gamma_{n,GaAs}$ where $\chi$ represents the leakage of the confined subband wavefunction into the AlGaAs (effectively modelling deviation from an infinite potential well). We estimated $\chi$ from simulations of the wavefunction of the lowest bound state in the potential wells in both confinement directions using the program developed by Gregory Snider for this purpose~\cite{snider}, exploring a rather large range of possible effective masses (0.2--0.65$m_0$, where $m_0$ is the bare electron mass), guided by theoretical values given in Reference~\cite{winkler} and the little that is experimentally known about the effective masses in these systems~\cite{nakwaski,manfra-densities,manfra-wellwidths}. We obtained $0 < \chi < 0.09$ and chose a value in the middle for the model. We note that $\chi$ is in any case a parameter to which our results are rather insensitive.

As noted above, $C$ in Reference~\cite{winkler}, which has a value of 0.009271, has been set to zero as it is in any case too small to perceptibly affect the results of the model.

The remaining four parameters $d_y$, $d_z$, $r_y$, $r_z$, are used as fitting parameters. They are grouped into two two-parameter sets which are treated differently; $d_y$ and $d_z$, the confinement related parameters are constrained to reasonable values --- not too different from an estimated extent of the wave function along these two directions, as explained below. We are uncertain about the values of $d_y$ and $d_z$ because while the thicknesses of the layers giving rise to the confinement are known to great accuracy, the actual extent of the wavefunctions and the precise relation of those to $d_y$ and $d_z$ are difficult to calculate. The other two parameters, $r_y$ and $r_z$, the Rashba coefficients are simply tuned to match the model calculations to the data. The values found are in fact the first available estimates of these parameters. This is because most work on SO in the valence band of GaAs has focused on heavy holes, whereas these are the parameters relevant for the light-hole-like subbands. As noted in the main article, we find the Rashba field to be much stronger in the $z$ direction than in the $y$ direction.

We have estimated the extent of the wavefunction in each direction rather crudely --- by modelling two fictitious two dimensional hole gases separately, neglecting the joint effect of both confinements in the real device. The two 2DHGs were simulated using nominal growth parameters in each direction separately. We found the extent of the wavefunction in $z$ to be about 15nm, while in $y$, we obtain about 30-35nm. The values for $d_y$ and $d_z$ used in the model are within a factor of two of these figures.

Finally, we note that our nonlinear transport measurements in the same device (to be published elsewhere) point to a subband spacing of about 3meV, which is slightly smaller than what the model predicts with the chosen parameters ($\sim$4meV, Figure 4a of the main article); we suspect that this mismatch is due to the mixing with nearby subbands (next few conductance steps), which are not considered in this model.

\subsection{Data at Intermediate Fields}

\begin{figure}
\flushleft
\includegraphics[0,0][60mm,160mm]{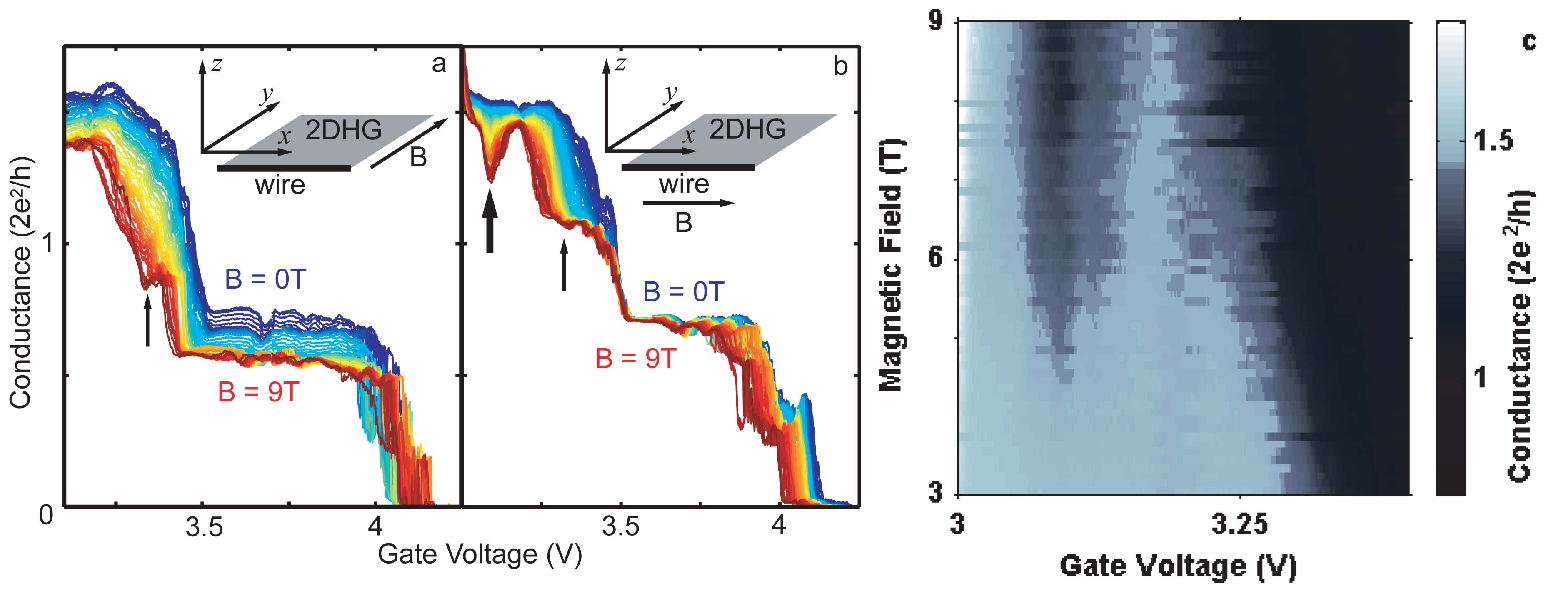}
\end{figure}

Figures a and b show data from Figure 3 in the main text, with additional traces taken at intermediate fields (every 0.2T) showing that the features discussed in the text (arrows) develop gradually with field and are not spurious effects. Figure c shows the data in Figure b around the spin-orbit gap.


\begin{thebibliography}{36}
\bibitem[{foo({\natexlab{a}})}]{footnote1}
\bibinfo{note}{Note that $\overrightarrow{B}_{SO}$ can be anywhere in the
  $y$-$z$ plane in Figure 2.}
\bibitem[{foo({\natexlab{b}})}]{footnote-stepheight}
\bibinfo{note}{Unlike in quantum point contacts, the conductance steps in these
  systems do not correspond to the ideal value of $2e^2/h$ due to the imperfect
  coupling between the 2DHG and the 1D wire as described in greater detail
  in~\cite{rafi-2d1d}.}
\bibitem[{foo({\natexlab{c}})}]{footnote-luttinger}
\bibinfo{note}{This is not to be confused with the Tomonaga-Luttinger liquid
  theory.}
\bibitem[{foo({\natexlab{d}})}]{footnote-smallstep}
\bibinfo{note}{The small half-step in the red trace in Figure 2d could be
  obscured in the data by the disorder-related features near depletion.}
\bibitem[{foo({\natexlab{e}})}]{footnote-gfactor}
\bibinfo{note}{One must be wary when speaking of effective g-factors in the
  presence of SO coupling as quantities commonly interpreted as
  `spin-splittings' are not necessarily equal to $gµ_BB$, where $g^*$ is the
  effective g-factor in the Zeeman term in the Hamiltonian. For example, in the
  simple 1D model described in Figure 2, the situation in Figures 2a and d
  could be interpreted as $g^* = 0$ based on measurements of subband edges such
  as those in References [25] and [26] whereas $g^*$ in the model is most
  definitely finite.}
\end{thebibliography}
\end{document}